\documentclass[aps,pre,showpacs,twocolumn]{revtex4-1}

\usepackage{amsmath,amssymb,amsfonts,hyperref}
\usepackage{graphicx,epsf}
\usepackage{xspace}
\usepackage{textcomp}
\usepackage{color}

\newif\ifhyper
\hypertrue
\ifhyper
\hypersetup{
  citecolor = {green},
  urlcolor = {blue} 
} 

\newcommand{\p}{\partial} 
\newcommand{\la}{\langle} 
\newcommand{\ra}{\rangle} 
\newcommand{\vx}{\vec{x}}
\newcommand{\vy}{\vec{y}}
\newcommand{\vz}{\vec{z}}
\newcommand{\vv}{\vec{v}}
\newcommand{\vu}{\vec{u}}
\newcommand{\vf}{\vec{f}\,}
\newcommand{\vJ}{\vec{J}}

\newcommand{\bx}{{\bf x}}
\newcommand{\by}{{\bf y}}
\newcommand{\bz}{{\bf z}}

\begin{document}

\title{Fully developed  isotropic turbulence: symmetries and exact identities}

\author{L\'eonie Canet$^{1,2}$, Bertrand Delamotte$^{3,4}$, and  Nicol\'as Wschebor$^{3,4,5}$}
\affiliation{
$^1$ Universit\'e Grenoble Alpes, UMR 5493, LPMMC, F-38000 Grenoble, France\\
$^2$  CNRS, UMR 5493, LPMMC, F-38000 Grenoble, France\\
$^3$ Sorbonne Universit\'es, UPMC Univ. Paris 06, UMR 7600, LPTMC, F-75005 Paris, France\\
$^4$ CNRS, UMR 7600, LPTMC, F-75005 Paris, France\\
$^5$Instituto de F\'isica, Facultad de Ingenier\'ia, Universidad de la Rep\'ublica, J.H.y Reissig 565, 11000 Montevideo, Uruguay}

\begin{abstract}

We consider the regime of fully developed  isotropic and homogeneous turbulence of the Navier-Stokes equation
 with a stochastic forcing.  We present two {\it gauge} symmetries of the corresponding Navier-Stokes field theory,
 and derive the associated general Ward identities. Furthermore, by introducing a local source bilinear in the velocity field, 
 we show that these symmetries entail an infinite set of {\it exact} and {\it local} relations between correlation functions.
 They include in particular the K\'arm\'an-Howarth relation and another exact relation  for a pressure-velocity correlation function
 recently derived  in Ref.~\cite{Falkovich10}, that we further generalize.

\end{abstract}

\maketitle

\section{Introduction}

The application of field-theoretic methods to Navier-Stokes (NS) turbulence has a long history \cite{Smith98,Zhou10}.
The most systematic way to implement them consists in deriving the generating functional  ${\cal Z}$ of velocity 
 correlation and response functions under the form of a functional integral, using the standard Martin-Siggia-Rose-Janssen-de Dominicis
 procedure (recalled below). The resulting action ${\cal S}$ involves not only the original fields  -- velocity $\vv(t,\vx)$ and pressure $p(t,\vx)$ -- but also the associated response fields $\bar \vv(t,\vx)$ and $\bar p(t,\vx)$. The  symmetries of the model and of the field theory 
 play a crucial role, especially regarding the renormalization properties of the model, since symmetries can yield  important  non-renormalization theorems. From this point of view, Kolmogorov exact result for the three-velocity correlation function should follow  from a symmetry of the NS field theory, but this is not the way it is usually presented. 

 In this work,  we perform a detailed analysis of the symmetries of the NS field theory. 
  We review the well-known Galilean invariance, and  the gauged (in time) version of this symmetry. Furthermore,  we   exhibit a gauged (in time)
  shift symmetry of the response velocity field, that was not, to the best of our knowledge, identified yet and which constitutes the first main contribution of this paper. Both these symmetries yield very useful Ward identities, that we derive. We show in Ref.~\cite{Canet14b} that, in the framework of 
  a non-perturbative renormalization group analysis,
  they eventually lead to a first-principles understanding of multi-scaling.

   Furthermore, by adding in the generating functional~${\cal Z}$ a source term bilinear in the velocity field  
$\exp(\int_{t,\vx} v_\alpha L_{\alpha\beta}v_\beta)$, we show that the response velocity shift symmetry can be 
fully gauged, both in time and space. The functional Ward identity ensuing 
 from this gauge symmetry yields infinitely many exact relations among correlation and response functions. 
  This constitutes the second main contribution of this paper.
 We show in particular that the K\'arm\'an-Howarth identity (which roots the  four-fifths Kolmogorov exact result for 
 the third order structure function),  emerges as a consequence of this gauge symmetry. This general 
 Ward identity also allows us to recover the exact relation, involving a pressure-velocity correlation function, derived in Ref~\cite{Falkovich10}, and to  generalize it.

   Let us briefly expound on the context of these symmetry studies. The Galilean invariance is a fundamental property of the theory of turbulence, and its implications have been widely  studied and discussed, in particular in field-theoretic descriptions. General Ward identities for time-independent Galilean 
 invariance were early derived, {\it e.g.} in Refs. \cite{dedominicis79,teodorovich89}. Time-dependent forms, also referred to as time-gauged, or extended, Galilean invariance
 were also introduced, {\it e.g.} in Refs. \cite{dedominicis79,adzhemyan88,adzhemyan94,antonov96}, mainly to derive exact results on the dimensions of composite operators in the operator product expansion.  Functional Ward identities associated with both the gauged and non-gauged Galilean symmetry were 
  also obtained in Ref. \cite{berera07,berera09} {\it via} the formalism of gauge fixing and Slavnov-Taylor identities related to the ensuing Becchi-Rouet-Stora (BRS) symmetry. 

  It is worth noting that a time-gauged Galilean symmetry was also unveiled in Ref.~\cite{Lebedev94} in a different but closely-related context:  in the field theory associated with the
 Kardar-Parisi-Zhang equation \cite{kardar86}, describing the stochastic growth and roughening of interfaces.
  This stochastic field theory shares 
 several common features with the NS field theory (for a detailed discussion of the corresponding Ward identities, 
 see  Refs.~\cite{Canet11a,Canet12Err}).  Besides the time-gauged Galilean invariance,  it is endowed with a
   time-gauged shift symmetry, highlighted in Ref. \cite{Canet11a},  reminiscent of the one we  bring out here for the NS field theory, although  it applies in the KPZ context to  the original field  instead of the response field.  
   Furthermore, as already mentioned, this symmetry admits, for the NS field theory, a fully gauged form provided  the bilinear source term
    $\vv\cdot L\cdot \vv$ is added, which yields more stringent Ward identities.

\section{Navier-Stokes equation with stochastic forcing}

To describe fully developed isotropic and homogeneous turbulence, one usually considers  the forced Navier-Stokes equation:
\begin{equation}
 \partial_t v_\alpha+ v_\beta \partial_\beta v_\alpha=-\frac 1\rho \partial_\alpha p +\nu \nabla^2 v_\alpha+f_\alpha
\label{ns}
\end{equation}
where the velocity field $\vv$, the pressure field $p$, and the forcing $\vec f$ depend on the 
space-time coordinates $(t,\vx)$, and with  $\nu$  the kinematic viscosity and $\rho$ the density of the fluid.
  The presence of a forcing $\vf$ is essential to balance
 the dissipative nature of the  (unforced) NS equation and maintain a turbulent steady state.
We consider in the following incompressible flows, satisfying 
\begin{equation}
 \partial_\alpha v_\alpha = 0.
\label{compress}
\end{equation}
Within the inertial range, correlation functions are expected to be universal and thus insensitive to the precise mechanism of   forcing at the integral scale $L$. One can hence conveniently average over various stochastic forcings $\vf$ with a Gaussian probability distribution, chosen  of zero mean and variance
\begin{equation}
 \langle f_\alpha(t,\vx)f_\beta(t',\vx\,')\rangle=2 \delta(t-t')N_{\alpha \beta}(|\vx-\vx\,'|)
\end{equation}
where the forcing  $N_{\alpha \beta}$ is exerted at the integral scale $L$.

\section{Navier-Stokes field theory}

The NS equation in the presence of a stochastic forcing $\vf$ stands as a Langevin equation. One can resort to the 
standard Martin-Siggia-Rose-Janssen-de Dominicis procedure \cite{martin73,Janssen76,DeDominicis76} to derive the associated field theory. 
It may be achieved in two different but equivalent ways.  Either the pressure field may be eliminated 
since the incompressibility constraint (\ref{compress}) uniquely fixes the pressure in terms of $\vf$ and $\vv$ 
as the solution of the  Poisson equation
\begin{equation}
 -\frac 1 \rho \partial_\alpha\partial_\alpha p=\partial_\alpha v_\beta \partial_\beta v_\alpha-\partial_\alpha f_\alpha
 \label{poisson}
\end{equation}
 for given boundary conditions or alternatively,  the pressure field may be kept and one introduces
  Martin-Siggia-Rose response fields $\bar v_\alpha$ and $\bar p$  to enforce both the equation of motion  (\ref{ns}) and the incompressibility constraint (\ref{compress}), which amounts in effect to encoding Eq.~(\ref{poisson}). We follow this latter route, because the pressure sector turns out
to be very simple to handle since it is  not renormalized (see Sec. \ref{SECpressure}). Moreover, this procedure ensures the analyticity of the effective action $\Gamma$, whereas  eliminating the pressure requires one to deal with non-analytic  (in wave-vectors) transverse projectors. As derived in Appendix A, once the response fields are introduced, the stochastic forcing can be integrated out
 and  one obtains the generating functional
\begin{equation}
 {\cal Z}[\vJ,\bar{\vJ},K,\bar{K}] = \int \mathcal{D}\vv \,\mathcal{D}p \,\mathcal{D}\bar{\vv} \,\mathcal{D}\bar p\,
  \, e^{-{\cal S}[\vv,\bar{\vv},p,\bar p]+{\cal J}[\vv,\bar{\vv},p,\bar p]} 
\label{Z}
 \end{equation}
where the  term ${\cal J}$ contains the
 sources $\vJ$,  $K$, $\bar{\vJ}$ and $\bar K$ for the velocity,  pressure and response fields
\begin{equation}
 {\cal J}[\vv,\bar{\vv},p,\bar p] = \int_{\bx}\big\{ \vJ\cdot \vv+\bar{\vJ}\cdot \bar{\vv}+K p+\bar K \bar p\big\}
\label{sources}
\end{equation}
with the notation $\bx \equiv (t,\vx)$ and $\int_\bx \equiv \int dt d^d \vx$, and
  where the NS action ${\cal S}$ is given by
\begin{align}
 {\cal S}[\vv,\bar{\vv},p,\bar p]&= \int_{\bx}\Big\{\bar p (\bx)\,\partial_\alpha v_\alpha(\bx)+\bar v_\alpha(\bx)\Big[ \partial_t v_\alpha(\bx)
 \nonumber\\
 &\hspace{-0.8cm}+v_\beta (\bx)\partial_\beta v_\alpha(\bx)+\frac 1\rho \partial_\alpha p (\bx)-\nu \nabla^2 v_\alpha(\bx) \Big]\Big\}\nonumber\\& -\int_{t,\vec x,\vec x'}\bar v_\alpha(t,\vec x) N_{\alpha \beta}(|\vec x-\vec x'|)\bar v_\beta(t,\vec x').
\label{NSaction}
\end{align}
We introduce for later purpose the notation $\varphi_i$, $i=1,\dots,4$, which stands for the fields $\vv,\bar \vv,p$ and $\bar p$, respectively,  and  $j_i$, $i=1,\dots,4$, which stands  for the sources $\vJ,\bar \vJ,K$ and $\bar K$, respectively.

 Field expectation values in the presence of the external sources $j_i$ 
 are obtained as functional derivatives of  ${\cal W} = \log {\cal Z}$ as 
\begin{equation}
  u_\alpha({\bf x}) = \langle v_\alpha({\bf x}) \rangle = \frac{\delta {\cal W}}{\delta J_\alpha({\bf x})}  \, \, , \, \,
  \bar u_\alpha({\bf x}) = \langle \bar v_\alpha({\bf x}) \rangle = \frac{\delta {\cal W}}{\delta \bar J_\alpha({\bf x})} \nonumber
\end{equation} 
 and similarly for the pressure fields, for which  for simplicity the same notation can be kept for the fields and their average values  
\begin{equation}
  p({\bf x}) \equiv \langle p({\bf x}) \rangle = \frac{\delta {\cal W}}{\delta K({\bf x})}  \, \, , \, \,
  \bar p({\bf x}) \equiv \langle \bar p({\bf x}) \rangle = \frac{\delta {\cal W}}{\delta \bar K({\bf x})} .\nonumber
\end{equation} 
The effective action $\Gamma[\vu,\bar\vu,p,\bar p]$ is defined as the Legendre transform of  ${\cal W}$:
\begin{equation}
\Gamma[\vu,\bar\vu,p,\bar p] +{\cal W}[\vJ,\bar{\vJ},K,\bar K] = {\cal J}[\la \varphi_i \ra].
\label{legendre}
\end{equation}

\section{Symmetries and related Ward identities}
\label{SYM}

In this section, we analyze  three gauge symmetries of the NS action (\ref{NSaction}), the time-gauged Galilean and response velocity  
field shift symmetries, and also gauged shifts of the pressure fields. More precisely, for each of these symmetries, the 
different terms of the NS action ${\cal S}$ are either invariant, or have a linear variation in the fields, and this entails
  important non-renormalization theorems and general Ward identities.
 The mechanism is standard: One
  performs as a change of variables in the functional integral ${\cal Z}$, Eq.~(\ref{Z}), the (infinitesimal) transformation 
  of the fields corresponding to the symmetry under study. Since it must leave ${\cal Z}$ unaltered, one  obtains the equality
 $\la \delta{\cal S}\ra =\la \delta{\cal J}\ra$. Then, because the variation of the action is linear in the fields, 
   $\la \delta{\cal S}[\varphi_i]\ra =  \delta{\cal S}[\la\varphi_i\ra]$
  and 
\begin{align}  
\la \delta{\cal J}\ra &= \la  \sum_i j_i \delta \varphi_i\ra  =  \sum_i \frac{\delta \Gamma}{\delta \la \varphi_i\ra} \la\delta \varphi_i\ra= \sum_i \frac{\delta \Gamma}{\delta\la\varphi_i\ra} \delta \la \varphi_i \ra\nonumber \\
 &\equiv \delta \Gamma
\end{align}
   using the definition (\ref{legendre}) of the Legendre transform.  The last two equalities hold because the considered 
   field transformations are all affine in the fields. The resulting identity hence imposes that the variation of the effective 
   action $\Gamma$ remains identical to the variation of the bare one $\cal S$ (in terms of the average fields). Thus, apart 
   from the linearly varying terms, 
 $\Gamma$ must be invariant under the considered symmetry, and
   these (non-symmetric) terms cannot be renormalized. These identities are now derived in detail for the three gauge symmetries.

\subsection{Non-renormalization of the pressure sector}
\label{SECpressure}

Let us begin with the simplest identities, the field equations for the pressure sector.
We consider the  infinitesimal field transformation   $p(t,\vx)\to p(t,\vx)+\epsilon(t,\vx)$ which describes a 
 gauged shift of the pressure field (and could be used to gauge it away  \footnote{The NS action is strictly invariant under a time-dependent shift  of the pressure $p(t,\vx)\to p(t,\vx)+\epsilon(t)$, as recalled {\it e.g.} in    
  Ref.~\cite{altaisky91}.}).
 This change of variables leaves the functional integral (\ref{Z})  unchanged, and thus one deduces
\begin{equation}
 0= \int_{\bx} \left\langle -\frac{\bar v_\alpha(\bx) \partial_\alpha \epsilon(\bx)}{\rho}+K(\bx)\epsilon(\bx)\right\rangle.
\end{equation}
Since this equality holds for all infinitesimal $\epsilon(t,\vx)$,  this yields
\begin{equation}
 K(\bx)=\frac{\delta \Gamma}{\delta p(\bx)}=-\frac{ \partial_\alpha \bar u_\alpha(\bx) }{\rho}=\frac{\delta S}{\delta p(\bx)}\Big|_{\varphi_i=\la\varphi_i\ra}\;\;,
\end{equation}
which means that the dependence in $p(t,\vx)$ of both  the effective action  $\Gamma$ and the bare one ${\cal S}$ are identical.
  A similar identity can be derived for the response pressure field $\bar p$, considering the infinitesimal gauged field transformation   $\bar p(t,\vec x)\to \bar p(t,\vec x)+\bar \epsilon(t,\vec x)$, which reads
\begin{equation}
 \bar K(\bx)=\frac{\delta \Gamma}{\delta \bar p(\bx)}=\p_\alpha u_\alpha(\bx)= \frac{\delta S}{\delta \bar p(\bx)}\Big|_{\varphi_i=\la\varphi_i\ra}\;\;. \label{Wshiftbarp}
\end{equation}
 One concludes that the whole pressure sector is not renormalized.

\subsection{Time-gauged Galilean symmetry}
\label{SYM-GAL}

We review in this section the time-gauged (or time-dependent) Galilean symmetry, and re-derive the corresponding   functional Ward identity.
 More precisely, the variation of the NS action (\ref{NSaction}) under a time-gauged  Galilean transformation 
 is linear in the fields, and 
 this entails a Ward identity, which is stronger than the usual non-gauged one.
Let us hence consider the infinitesimal time-gauged Galilean transformation ${\cal G}(\vec\epsilon(t))$, defined as
\begin{align}
 \delta v_\alpha(\bx)&=-\dot{\epsilon}_\alpha(t)+\epsilon_\beta(t) \partial_\beta v_\alpha(\bx)\nonumber\\
 \delta \bar v_\alpha(\bx)&=\epsilon_\beta(t) \partial_\beta \bar v_\alpha(\bx)\nonumber\\
 \delta p(\bx)&=\epsilon_\beta(t) \partial_\beta p(\bx)\nonumber\\
 \delta \bar p(\bx)&=\epsilon_\beta(t) \partial_\beta \bar p(\bx)
\label{defG}
 \end{align}
where $\dot{\epsilon}_\alpha = \p_t \epsilon_\alpha$.
The NS action is invariant under ${\cal G}(\vec\epsilon)$ with an arbitrary constant vector $\vec \epsilon\,$, which corresponds to a  
translation in space, and also under  ${\cal G}(\vec\epsilon\, t)$, which corresponds to the usual (non-gauged) Galilean transformation.

To analyze the variation of the NS action under a generic  transformation ${\cal G}(\vec\epsilon(t))$, let us  define,
 following the standard geometric interpretation  \cite{antonov96},
 a Galilean scalar density, as a quantity $\psi(\bx)$ which transforms under (\ref{defG}) 
 as $\delta \psi(\bx)=\epsilon_\beta(t) \partial_\beta \psi(\bx)$.  This definition implies that the integral 
 over $\vx$ of a scalar density is invariant under ${\cal G}$ since $\int_{\vx} \delta\psi = 0$.  
 The sum and the product of two scalar densities are  scalar densities.  
 The gradient of a scalar density is also a scalar density, whereas its time derivative is not. 
 However,  the Lagrangian time derivative, defined by
\begin{equation}
  D_t \psi(\bx)\equiv  \partial_t \psi(\bx)+ v_\beta(\bx) \partial_\beta \psi(\bx),
\end{equation}
is a covariant time derivative since   $D_t \psi$ remains a scalar  density if $\psi$ is. 

 The three fields $\bar{\vv}$, $p$ and $\bar p$ are scalar densities by definition of the transformation 
 (\ref{defG}), and so is the gradient of the velocity field $\vv$. In contrast, neither $\vv$ nor $\p_t \vv$ are scalar  densities. 
 One concludes that all the terms in the NS action (\ref{NSaction}) are invariant under the time-gauged 
 transformation ${\cal G}(\vec\epsilon(t))$, apart from the term proportional to the Lagrangian time derivative of the velocity 
 $D_t v_\alpha(\bx)$. 
 Nevertheless, although this is not strictly a scalar density as its transform includes an additional contribution proportional to $ \ddot{\epsilon}_\alpha(t)$, the variation of
  the corresponding term in the action is linear in the fields. In summary, the overall variation of the action ${\cal S}$ is 
\begin{equation}
 \delta {\cal S}= \delta \int_{\bx} \bar v_\alpha(\bx) D_t  v_\alpha(\bx)=
 -\int_{\bx} {\epsilon}_\alpha(t) \p_t^2 \bar v_\alpha(\bx).
\end{equation}
 Hence, performing  the transformation ${\cal G}(\vec\epsilon(t))$ in the functional integral  (\ref{Z}), one obtains
\begin{align}
 \la \delta {\cal S}\ra &=  \la \delta {\cal J}\ra = \int_\bx \Big\{ -\dot\epsilon_\alpha(t)  J_\alpha(\bx) + \epsilon_\beta(t) \sum_i j_i  \p_\beta \la \varphi_i \ra\Big\} \nonumber \\
 & = \delta \Gamma
\end{align}
which, since the equality is valid for all infinitesimal $\vec\epsilon(t)$,
 yields the following Ward identity
\begin{align}
\label{wardgalilee}
&  \int_{\vx} \Big\{\big(\delta_{\alpha\beta}\p_t +  \partial_\alpha u_\beta(\bx)\big)\frac{\delta \Gamma}{\delta u_\beta(\bx)}   \nonumber
+ \partial_\alpha \bar u_\beta(\bx) \frac{\delta \Gamma}{\delta \bar u_\beta(\bx)} \nonumber\\
 &+\partial_\alpha p(\bx) \frac{\delta \Gamma}{\delta p(\bx)}  
 +\partial_\alpha \bar p(\bx) \frac{\delta \Gamma}{\delta \bar p(\bx)} \Big\}= -\int_{\vx} \p_t^2 \bar u_\alpha(\bx).
\end{align}
Thus, both  variations of the effective action and  of the bare one are identical.
This entails that, apart from the term $\int_{\bx}\bar u_\alpha(\bx) D_t u_\alpha(\bx)$ which is not renormalized and remains equal to its bare expression,
$\Gamma$ is invariant under time-gauged Galilean transformations.

\subsection{Time-gauged response fields shift symmetry}

Let us show that another class of transformations yields non-renormalization theorems because again the corresponding variation of the NS action ${\cal S}$ is linear in the fields. They consist in time-gauged shifts of the response fields, represented by  the infinitesimal transformation  \footnote{The NS action is strictly invariant under a constant shift of the velocity response field, as noted  {\it e.g.} in \cite{teodorovich89}.}
\begin{align}
\label{shiftjauge}
 \delta \bar v_\alpha(\bx)&=\bar \epsilon_\alpha(t) \nonumber\\
 \delta \bar p(\bx)&= v_\beta(\bx) \bar \epsilon_\beta(t).
 \end{align}
 This transformation leaves in particular the combination $\bar v_\alpha v_\beta \partial_\beta v_\alpha+\bar p \partial_\alpha v_\alpha$ invariant. The overall variation of the action stems from the term $\int_\bx \bar v_\alpha \p_t v_\alpha$ and from the forcing term and is  linear in the fields:
\begin{equation}
 \delta {\cal S} = \int_{\bx}  {\bar\epsilon}_\beta(t) \p_t v_\beta(\bx) +2 \int_{t,\vx,\vx'} {\bar\epsilon}_\alpha(t) N_{\alpha_\beta} (\vx-\vx') \bar v_\beta(t,\vx').
\end{equation}
 Hence, performing the change of variables (\ref{shiftjauge}) in  the functional integral  (\ref{Z}), one concludes that the variations of both the bare and effective actions coincide, and deduces the following Ward identity
 \begin{align}
\label{wardshift}
\int_{\vx} \Big\{\frac{\delta \Gamma}{\delta \bar u_\alpha(\bx)}
 &+ u_\alpha(\bx) \frac{\delta \Gamma}{\delta \bar p(\bx)} \Big\}= \int_{\vx}  \p_t u_\alpha(\bx) \nonumber\\
 &+ 2 \int_{\vx,\vx'}N_{\alpha_\beta} (\vx-\vx') \bar u_\beta(t,\vx') .
\end{align}
Again, this implies that, apart from the term $\int_{\bx} \bar u_\alpha \partial_t u_\alpha$ and the forcing term that are not 
renormalized, the effective action $\Gamma$ is invariant under time-gauged response fields shift transformations.

\subsection{General structure of the effective action $\Gamma$}

Let us  summarize  the previous analysis of the symmetries of the NS field theory. The effective action $\Gamma$ may be written as
\begin{align}
\Gamma[\vu,\bar \vu,& p,\bar p] = \int_{\bx}\Big\{ \bar u_\alpha \Big(\partial_t u_\alpha+  u_\beta \partial_\beta u_\alpha +\frac{\partial_\alpha p}{\rho}\Big)+\bar p \partial_\alpha u_\alpha\Big\}\nonumber\\
& -\int_{t,\vx,\vx'}\Big\{\bar u_\alpha(t,\vx) N_{\alpha\beta}(\vx-\vx') \bar u_\beta(t,\vx') \Big\}  +\tilde \Gamma[\vu,\bar \vu]
 \label{anzGk}
\end{align}
where the explicit terms are not renormalized and thus keep their bare forms,
  and the functional  $\tilde \Gamma$ only depends on the velocity fields and is invariant under time-gauged Galilean and  response velocity shift transformations.

\section{Exact relations in the presence of a local bilinear source}

In this section we consider the same model in the presence of a local source $L_{\alpha\beta}$ for the quadratic operator
$v_\alpha(\vec x,t) v_\beta(\vec x,t)$. The advantage of adding such a source is that the response fields
 shift symmetry (\ref{shiftjauge}) can be  completely  gauged (both in time {\it  and space}). 
  The ensuing functional Ward identity yields in particular the well-known exact K\'arm\'an-Howarth relation \cite{karman38},
   from which can be derived the four-fifths Kolmogorov law for the $S^{(3)}$
structure function \cite{Frisch95}.  This Ward identity also entails the exact relation for a pressure-velocity correlation function   
derived in \cite{Falkovich10}, and further allows one to extend it.
More generally, it constitutes a full functional relation from which can be deduced an infinite set of
exact local relations which, to the best of our knowledge, were not obtained before. We first derive this Ward identity for the
effective action $\Gamma$. However, since in the literature the known relations are expressed
 in terms of the connected correlation functions,  we also formulate this functional relation for the
corresponding generating functional. 

\subsection{Local Ward identity for the effective action}

In order to deduce Ward identities, we consider the generalized generating functional in the presence of a local 
source $L_{\alpha\beta}(\bx)$ for the composite operator $v_{\alpha}(\bx)v_{\beta}(\bx)$
\begin{equation}
 {\cal Z}[\vJ,\bar{\vJ},K,\bar K,L] = \int \mathcal{D}\vv \,\mathcal{D}p \,\mathcal{D}\bar{\vv} \,\mathcal{D}\bar p\,
  \, e^{-{\cal S}[\vv,\bar{\vv},p,\bar p] + {\cal J}_L[\vv,\bar{\vv},p,\bar p]}
\label{fonctgeneratgeneral}
 \end{equation}
with the new source term
\begin{equation}
 {\cal J}_L[\vv,\bar{\vv},p,\bar p] = \int_{\bx}\{ \vJ\cdot \vv+\bar{\vJ}\cdot \bar{\vv}+K p+\bar K\bar p+\vv\cdot L \cdot \vv\}
\end{equation}
 using  matrix notation $\vv\cdot L \cdot \vv\equiv v_\alpha(\bx) L_{\alpha\beta}(\bx)  v_\beta(\bx)$. 
 We now consider a shift in the response fields gauged both in space and time:
\begin{align}
\label{shiftjaugeespettemps}
 \delta \bar v_\alpha(\bx)&=\bar \epsilon_\alpha(t,\vec x), \nonumber\\
 \delta \bar p(\bx)&= v_\beta(\bx) \bar \epsilon_\beta(t,\vec x).
 \end{align}
None of the terms of the NS action (\ref{NSaction}) are invariant under this transformation, but their variations
 are still linear in the fields. Thus, using the same mechanism, one can derive Ward identities, which are now completely local,
  {\it i.e.} no longer integrated over space, since the relation $\la\delta {\cal S}\ra = \la \delta {\cal J}_L\ra$ obtained when 
 performing the change of variables (\ref{shiftjaugeespettemps}) in 
 the generating functional (\ref{fonctgeneratgeneral}) is now valid for all $\bar {\vec\epsilon}(t,\vx)$. 
  One obtains
 \begin{align}
 \label{Wardtempsetesp}
&\Big\langle-\partial_t v_\alpha-\frac{1}{\rho}\partial_\alpha p +\nu \nabla^2 v_\alpha -\partial_\beta(v_\alpha v_\beta)+\bar J_\alpha + \bar K v_\alpha \nonumber\\
&+\int_{\vec x'}\Big(2N_{\alpha\beta}(\vec x -\vec x') \bar v_\beta(t,\vec x')\Big)\Big\rangle = 0.
 \end{align}
The key role of the new source term is that the average $\langle\partial_\beta(v_\alpha v_\beta)\rangle$ can now be expressed 
as a derivative with
respect to $L_{\alpha\beta}$. The precise form of the generalized Legendre transform in the presence of this source is the 
same as without it:
\begin{align}
&  \Gamma[\vu,\bar \vu, p,\bar p,L]+{\cal W}[\vJ,\bar \vJ,K,\bar K,L] \nonumber\\
 &= \int_{\bx}\Big( \vJ \cdot \vu + \bar \vJ\cdot \bar \vu + K p+ \bar K \bar p \Big),
\end{align}
that is, (as usual with composite operators), we {\it do not perform} the Legendre transform with respect to the corresponding source.
It follows that  $\la v_\alpha v_\beta\ra = \frac{\delta {\cal W}}{\delta L_{\alpha\beta}} = - \frac{\delta \Gamma}{\delta L_{\alpha\beta}}$. Moreover, the Ward identities ensuing from the gauged shifts of the pressure fields are unchanged  in the presence of the  source $L$. 
Hence, using the explicit form of $\frac{\delta\Gamma}{\delta \bar p}$, Eq.~(\ref{Wshiftbarp}), the identity (\ref{Wardtempsetesp}) reads
 \begin{align}
\frac{\delta \Gamma}{\delta \bar u_\alpha} &=\partial_t u_\alpha+\frac{1}{\rho}\partial_\alpha p -\nu \nabla^2 u_\alpha
-\partial_\beta\Big(\frac{\delta \Gamma}{\delta L_{\alpha\beta}} \Big)
-u_\alpha\partial_\beta u_\beta\nonumber\\
&-2 \int_{\vec x'} N_{\alpha\beta}(\vec x -\vec x') \bar u_\beta(t,\vec x').
\end{align}
This functional identity has several important properties.
As already emphasized, this identity  is {\it local} in space and time.
Moreover, it implies that vertex functions involving the response field $\bar \vu$ (that is, generalized response functions)
can be expressed in terms of vertex functions involving the composite operator $v_\alpha v_\beta$ in a simple way. More precisely, it
 means that $\Gamma[\vu,\bar \vu, p,\bar p,L]$ does not depend on $L$ and $\bar \vu$ independently.
Furthermore, let us point out that  the parameter $\nu$ explicitly enters this identity whereas it was not present
in the previous one (\ref{wardshift}) (integrated on space).
Let us finally mention that the Ward identity for $\Gamma$ stemming from the Galilean symmetry in the presence of the source $L$ keeps the same form as without it, except for the source  $L$ which  is added to the fields transforming as Galilean scalar densities (see below for ${\cal W}$).

\subsection{Local Ward identity for connected correlation functions}

Let us now write the same Ward identity for the generating functional
 of connected correlation functions $\cal{W}$.
 Eq.~(\ref{Wardtempsetesp}) can alternatively be expressed as  the following Ward identity:
\begin{align}
-\partial_t \frac{\delta \cal{W}}{\delta J_\alpha}&-\frac{1}{\rho}\partial_\alpha\frac{\delta \cal{W}}{\delta K}
+\nu \nabla^2 \frac{\delta \cal{W}}{\delta J_\alpha}+ \bar J_\alpha + \bar K \frac{\delta \cal{W}}{\delta J_\alpha} 
-\partial_\beta \frac{\delta \cal{W}}{\delta L_{\alpha\beta}}\nonumber\\
&+  \int_{\vec x'} \Big\{ 2 \frac{\delta \cal{W}}{\delta \bar J_\beta(t,\vec x')} N_{\alpha\beta}(\vec x-\vec x') \Big\}=0  .
\label{wardW}
\end{align}

For completeness, we  derive the time-gauged  Galilean Ward identity for ${\cal W}$ in the presence of $L$. As previously, performing the
change of variables (\ref{defG}) in the generating functional  (\ref{fonctgeneratgeneral}) yields the Ward identity:
\begin{align}
 \int_{\vec x} \Big\{  \partial_t^2 \frac{\delta {\cal W}}{\delta \bar J_\alpha}& +\partial_t J_\alpha
 +  J_\beta \partial_\alpha \frac{\delta {\cal W}}{\delta J_\beta}+ K \p_\alpha\frac{\delta {\cal W}}{\delta p}+ \bar K \p_\alpha\frac{\delta {\cal W}}{\delta \bar p}\nonumber\\
 & +2 \p_t\Big(L_{\alpha\beta}\frac{\delta {\cal W}}{\delta J_\beta}\Big)
 +L_{\beta\gamma}\partial_\alpha\Big(\frac{\delta {\cal W}}{\delta L_{\beta\gamma}}\Big)\Big\}  =0.
\end{align}
This identity is very similar to the previous Ward identity (\ref{wardgalilee}) (but formulated here in terms of ${\cal W}$) except  that  $L$ has to be also considered as a scalar density with respect to Galilean transformations.

\subsection{K\'arm\'an-Howarth relation}

We now show that the Ward identity (\ref{wardW}) entails the  K\'arm\'an-Howarth relation. 
 For this, we differentiate this relation with
respect to $J_\beta(t_y,\vec y)$ and evaluate the resulting identity at zero external sources. This yields
\begin{align} 
&0=(\nu \Delta_x-\p_{t_x}) \la v_\alpha(\bx) v_\beta(\by)\ra -\frac 1 \rho \p^x_\alpha  \la p(\bx) v_\beta(\by)\ra \nonumber\\  &-\p_\gamma^x \la v_\alpha(\bx) v_\gamma(\bx) v_\beta(\by)\ra+ 2 \int_{\vx'}  \la \bar v_\gamma(\bx) v_\beta(\by)\ra N_{\alpha\gamma}(\vec x-\vec x') .
\end{align}
As shown  in Appendix A, the term involving $N_{\alpha\gamma}$
 can be simply expressed in term of the force as $\la f_\gamma(\bx) v_\beta(\by) \ra$.
 We now set $\alpha=\beta$ and sum over $\alpha$. By homogeneity, $\p^x_\alpha = -\p^y_\alpha$ when acting on averages, 
 and thus the term proportional to the pressure vanishes due to the incompressibility constraint \footnote{Note that the velocity field $v_\alpha$ is unconstrained in the functional integral (\ref{fonctgeneratgeneral}). However, since the  response pressure sector is not renormalized, the incompressibility constraint is in practice always satisfied on average at zero external sources.}.
 Symmetrizing in $\bx$ and $\by$, and considering equal times $t_x=t_y\equiv t$ eventually yields 
\begin{align}
&-\p_t \la v_\alpha(\bx) v_\alpha(\by)\ra +  \nu (\Delta_x+\Delta_y) \la v_\alpha(\bx) v_\alpha(\by)\ra \nonumber \\
& -\p_\gamma^x \la v_\alpha(\bx) v_\gamma(\bx) v_\alpha(\by)\ra -\p_\gamma^y \la v_\alpha(\by) v_\gamma(\by) v_\alpha(\bx)\ra\nonumber \\
& + \la f_\alpha(\bx) v_\alpha(\by) \ra+ \la f_\alpha(\by) v_\alpha(\bx) \ra=0,
\end{align}
 which identifies with  the K\'arm\'an-Howarth relation \cite{karman38}. The four-fifths Kolmogorov law \cite{Kolmogorov41c} 
 stating  that the third order structure function exactly obeys the Kolmogorov scaling follows from this relation \cite{Frisch95}.

 \subsection{Relation for the pressure-velocity correlation function of Ref.~\cite{Falkovich10}}

We now show that the Ward identity (\ref{wardW}) also yields another exact relation, recently derived in Ref. \cite{Falkovich10}.
 For this, let us differentiate twice Eq.~(\ref{wardW}) with respect to $L_{\mu\nu}(t_y,\vy)$ and $J_\beta(t_z,\vz)$
 and evaluate at zero external sources, to obtain
\begin{align}
 0&= (\nu \Delta^x -\p_{t_x})\la v_\alpha(\bx) v_\mu(\by) v_\nu(\by) v_\beta(\bz)\ra\nonumber\\
&-\frac 1 \rho \la p(\bx)v_\mu(\by) v_\nu(\by) v_\beta(\bz) \ra  +\la f_\alpha(\bx) v_\mu(\by) v_\nu(\by) v_\beta(\bz) \ra\nonumber\\
& -\p_\gamma^x \la v_\alpha(\bx)v_\gamma(\bx) v_\mu(\by) v_\nu(\by) v_\beta(\bz) \ra
\label{ward-Falkovich}
\end{align}
 where the term proportional to $N_{\alpha\beta}$ was expressed in term of the forcing $\vf$ following  Appendix A. 
 We now set $\mu=\nu$ and $\alpha=\beta$, sum over $\mu$ and $\alpha$, and eventually choose coinciding 
 space points $\vx=\vz$  and coinciding times $t_x=t_y=t_z$, which gives
 \begin{align}
 &\la  v_\alpha(\bx) v^2(\by)[\nu \Delta^x - \p_{t_x}] v_\alpha(\bx)\ra - \frac 1 \rho \la  v^2(\by) v_\alpha(\bx) \p^x_\alpha  p(\bx)\ra\nonumber\\
&  +\la f_\alpha(\bx) v_\alpha(\bx) v^2(\by) \ra - \la v_\alpha(\bx) v_\gamma(\bx)v^2(\by) \p^x_\gamma v_\alpha(\bx)\ra   =0.
\label{Falkovich-relation}
\end{align}
Notice that the correlation functions in this relation are connected since they all originate in differentiating the $\cal W$ functional.
However, one can easily show that the non-connected parts either vanish or factorize into $\la \vec{v}^2\ra$ times the left 
hand side of the  K\'arm\'an-Howarth relation, which hence also vanishes. Thus, Eq.~(\ref{Falkovich-relation}) holds for both
the connected and non-connected correlation functions.
 In the stationary regime, the term involving the time derivative is proportional to $\p_{t_x}\la v^2(\bx)v^2(\by)\ra$ and thus vanishes, and one obtains
\begin{align}
&  \nu \la v_\alpha(\bx) \Delta^x v_\alpha(\bx) v^2(\by) \ra -\frac 1 \rho \p_\alpha^x \la v^2(\by) v_\alpha(\bx) p(\bx)\ra  \nonumber\\
&+ \la f_\alpha(\bx)  v_\alpha(\bx) v^2(\by) \ra  -\frac 1 2  \p^x_\alpha \la v_\alpha(\bx)  v^2(\bx) v^2(\by) \ra =0 
 \label{relFalko}
\end{align}
using the incompressibility constraint [23]. Equation~(\ref{relFalko})
 coincides with the exact relation for the pressure-velocity correlation function derived in Ref. \cite{Falkovich10}.
 According to the authors of Ref. \cite{Falkovich10}, this relation can be  simplified in the inertial range
  using  arguments based on the decoupling of the large- and small-scale fields as:
  \begin{equation}
   \la \vec{v}(\vec{r}) p(\vec{r}) v^2(0)\propto \vec{r}.
  \end{equation}
  
  Other exact relations among (connected or non-connected) correlation functions
  involving one pressure field and three velocity fields can be simply generated  from
   Eq.~(\ref{ward-Falkovich}). For instance, setting $\mu=\alpha$ and $\nu=\beta$, summing over $\mu$ and $\nu$, 
  and eventually choosing coinciding 
 space points $\vx=\vz$  and coinciding times $t_x=t_y=t_z$ yields
 \begin{align}
 &0=\left\la  v_\alpha(\by) \vec{v}(\by).\vec{v}(\bx)\left[ (\nu \Delta^x - 
 \p_{t_x}) v_\alpha(\bx)-\frac 1 \rho \p_{\alpha}^x p(\bx)\right]\right\ra 
 \nonumber\\
&  +\la f(\bx).v(\by)\vec{v}(\by).\vec{v}(\bx)  \ra - 
\la v_\alpha(\by)\vec{v}(\by).\vec{v}(\bx) v_\gamma(\bx) \p^x_\gamma v_\alpha(\bx)\ra  .
\label{Falkovich-relation-2}
\end{align}
Using the same arguments put forwards in Ref.~\cite{Falkovich10},  
this relation can also be simplified in the stationary regime and in the inertial range, as detailed in Appendix B. 
We find:
\begin{align}
 \frac 1 \rho  \la \p_{\alpha}^x p(\bx) v_\alpha(0) &\vec{v}(0).\vec{v}(\bx)\ra
 =- \frac{d+2}{2d} \la \epsilon \vec{v}^2\ra+
 \la \vec{f}.\vec{v}\, \vec{v}^2 \ra \nonumber \\
 &-\frac{d}{64} F(x^2)-\frac{1}{32} F'(x^2)x^2
 \label{new-relation}
\end{align}
where
\begin{equation}
\la  v_\alpha(\bx) (\vec{v}(0).\vec{v}(\bx))^2 \ra = \left(\frac{1}{32} F(x^2) + \frac 1 d \la \epsilon \vec{v}^2 \ra\right) x_\alpha.
\end{equation}
Thus, one concludes that if $F$  does not diverge when $x\to 0$,  then Eq.~(\ref{new-relation}) implies that at leading order in $\vert \vec{x}\vert$,
$\la\p_{\alpha}^x p(\bx) v_\alpha(0)\vec{v}(0).\vec{v}(\bx)\ra$ is a constant, whereas if $F$ diverges, then 
 it follows that
\begin{equation}
\frac 1 \rho  \la \p_{\alpha}^x p(\bx) v_\alpha(0)\vec{v}(0).\vec{v}(\bx)\ra \simeq 
-\frac 1 2 \p_{\alpha}^x\la  v_\alpha(\bx) (\vec{v}(0).\vec{v}(\bx))^2 \ra .
\label{rel-F-2}
\end{equation}

\section{Conclusion}

In this paper, we  analyze  the symmetries of the NS field theory. 
 We revisit  in particular  the time-gauged form of the Galilean symmetry and
 unveil a time-gauged  response velocity shift symmetry, which, to the best of our knowledge, was not yet identified.
  We derive the 
  related general (functional) Ward identities.
 Furthermore, we show, by introducing a local  source term  bilinear in the velocity, that 
 the related Ward identities  yield an infinite set of {\it exact} and {\it local} relations between correlation functions.   They include in particular
 the K\'arm\'an-Howarth relation and a similar identity for the pressure-velocity correlation function recently derived  in Ref. \cite{Falkovich10}. Furthermore, we generalize this latter identity.

\section{Acknowledgments}

The authors gratefully thank M. Brachet for fruitful dicussions and for pointing out  several valuable references and G. Falkovich for  enlightening correspondence
about small and large scale fields.
  The authors acknowledge financial support from Ecos-Sud France-Uruguay program (code of grant U11E01) and from the Proyecto para el Desarrollo de las Ciencias Básicas (PEDECIBA, Uruguay). L.C. and B.D.  also 
  thank the Universidad de la Rep\'ublica (Uruguay) for hospitality during the completion of this work, and N.W. thanks the LPTMC for hospitality. 
 
\begin{appendix}

\renewcommand{\theequation}{A\arabic{equation}}
\section*{Appendix A: Field theory for the NS equation}
\setcounter{equation}{0}

In this Appendix, we recall the derivation of the field theory associated with the NS equation in the presence 
 of a stochastic stirring force $\vf$, following the standard Janssen-de Dominicis procedure \cite{Janssen76,DeDominicis76}. 
 Let us consider  the mean value of an observable ${\cal O}$, which is a functional of the velocity field, with respect to the distribution of
 the stochastic force. It is given by
\begin{equation}
 \langle \mathcal{O}[\vv]\rangle \propto \int \mathcal{D}\vf \mathcal{O}[\vv_f] \mathcal{P}[\vf]
\label{mean}
\end{equation}
where $\vv_f$ denotes the incompressible solution of the NS equation for the specific realization $\vf$ of the force (for fixed initial conditions, which are neglected as they are assumed to play no role in the stationary regime). The normalizations -- proportionality constants -- need not be specified either.
 The probability distribution $\mathcal{P}[\vf]$ of the force is Gaussian and assumed to be given by
\begin{equation}
\mathcal{P}[\vf]\propto\exp\Bigg(-\frac 1 4 \int_{t,\vec x,\vec x'}f_\alpha(t,\vec x) (N^{-1})_{\alpha \beta}(|\vec x-\vec x'|)f_\beta(t,\vec x')\Bigg).
\end{equation}
The mean value (\ref{mean}) can  be expressed as a functional integral over the velocity and pressure fields as
\begin{align}
 \langle \mathcal{O}[\vv]\rangle& \propto \int \mathcal{D}\vf \,\mathcal{D}\vv \,\mathcal{O}[\vv] \mathcal{P}[\vf] \delta(\vv-\vv_f)\nonumber\\
& \propto \int \mathcal{D}\vf\, \mathcal{D}\vv\, \mathcal{D}p\,
 \mathcal{O}[\vv] \mathcal{P}[\vf]\,\delta(\partial_\alpha v_\alpha)\times \delta({\cal E}_\alpha)
\label{meanv}
\end{align}
with 
\begin{equation}
 {\cal E}_\alpha \equiv \partial_t v_\alpha+ v_\beta \partial_\beta v_\alpha+\frac 1\rho \partial_\alpha p -\nu \nabla^2 v_\alpha-f_\alpha
\end{equation}
where the second line of (\ref{meanv}) holds in It$\bar{\rm o}$'s discretization for which  the transformation has a constant Jacobian independent of the fields.
 One  introduces the Lagrange multipliers
 $\tilde{\vv}$ and $\tilde p$ to enforce the equation of motion and the incompressibility constraint in the following way  
\begin{align}
 \langle \mathcal{O}[\vv]\rangle
& \propto \int \mathcal{D}\vf \,\mathcal{D}\vv \,\mathcal{D}p\, \mathcal{D}\tilde{\vv} \,\mathcal{D}\tilde p\,
 \mathcal{O}[\vv] \mathcal{P}[\vf] \nonumber\\
&\times \exp \Big\{-i\int_{\bx} \tilde v_\alpha \,{\cal E}_\alpha -i\int_{\bx}\tilde p\, \partial_\alpha v_\alpha\Big\} .
\end{align}
The functional integral on the force $\vf$ can then be carried out and one obtains
\begin{equation}
 \langle \mathcal{O}[\vv]\rangle \propto \int \mathcal{D}\vv \,\mathcal{D}p \,\mathcal{D}\tilde{\vv} \,\mathcal{D}\tilde p\,
 \mathcal{O}[\vv] \, e^{-{\cal S}[\vv,\tilde{\vv},p,\tilde p]}
 \end{equation}
with the action
\begin{align}
 {\cal S}[\vv,\tilde{\vv},p,\tilde p]&=i\int_{\bx}\Bigg\{\tilde p \,\partial_\alpha v_\alpha\nonumber\\
& +\tilde v_\alpha\Big[ \partial_t v_\alpha+ v_\beta \partial_\beta v_\alpha+\frac 1\rho \partial_\alpha p -\nu \nabla^2 v_\alpha \Big]\Bigg\}\nonumber\\
&+\int_{t,\vec x,\vec x'}\tilde v_\alpha(t,\vec x) N_{\alpha \beta}(|\vec x-\vec x'|)\tilde v_\beta(t,\vec x').
\end{align}
We finally redefine the fields as $\bar p \equiv i \tilde p$ and $\bar{\vu} \equiv i \tilde {\vu}$, and introduce sources $\vJ$, $K$, $\bar{\vJ}$, and $\bar K$ for the velocity, the pressure and the response fields, respectively. One can hence write the Janssen-de Dominicis generating functional as
\begin{align}
 {\cal Z}[\vJ,\bar{\vJ},K,\bar K] &= \int \mathcal{D}\vv \,\mathcal{D}p \,\mathcal{D}\bar{\vv} \,\mathcal{D}\bar p\,
  \, e^{-{\cal S}[\vv,\bar{\vv},p,\bar p] + {\cal J}[\vv,\bar{\vv},p,\bar p]} 
 \end{align}
 where ${\cal S}$ and ${\cal J}$ are given by Eqs.~(\ref{NSaction}) and (\ref{sources}), respectively.

Let us now calculate the mean value of a quantity which  depends linearly on the forcing:
\begin{equation}
 \langle f_\alpha(t,\vec x) {\cal O}[\vv] \rangle \propto \int \mathcal{D}\vec f\, \mathcal{P}[\vec f]\, f_\alpha(t,\vec x)  {\cal O}[\vv_f]
\end{equation}
with the same normalization factor as in Eq.~(\ref{mean}).
The Janssen-de Dominicis procedure can be applied almost in the same way yielding  
\begin{align}
 \langle f_\alpha(t,\vec x)&  {\cal O}[\vv]\rangle
\propto \int \mathcal{D}\vf \,\mathcal{D}\vv \,\mathcal{D}p\, \mathcal{D}\tilde{\vv} \,\mathcal{D}\tilde p\,
 \, f_\alpha(t,\vec x)  {\cal O}[\vv] \nonumber\\
&\times  \mathcal{P}[\vf] \exp \Big\{-i\int_{\bx} \tilde v_\alpha \,{\cal E}_\alpha -i\int_{\bx}\tilde p\, \partial_\alpha v_\alpha\Big\}.
\end{align}
One can now perform the Gaussian integral, taking into account that the quantity to be averaged depends on $f_\alpha$, to obtain
\begin{equation}
 \langle f_\alpha(t,\vec x)  {\cal O}[\vv]\rangle=2  \int_{\vec x'} N_{\alpha\beta}(|\vec x-\vec x'|) \langle  \bar v_\beta(t,\vec x')    {\cal O}[\vv]\rangle.
\end{equation}
The averages of quantities linear in $\vf$ are thus related to response functions. 

\renewcommand{\theequation}{B\arabic{equation}}
\section*{Appendix B: Simplification of Eq.~(\ref{Falkovich-relation-2})}
\setcounter{equation}{0}
We consider Eq.~(\ref{Falkovich-relation-2}), which reads:
\begin{align}
 &0=\left\la  v_\alpha(\by) \vec{v}(\by).\vec{v}(\bx)\left[ (\nu \Delta^x - 
 \p_{t_x}) v_\alpha(\bx)-\frac 1 \rho \p_{\alpha}^x p(\bx)\right]\right\ra 
 \nonumber\\
&  +\la \vec{f}(\bx).\vec{v}(\by)\vec{v}(\by).\vec{v}(\bx)  \ra - 
\la v_\alpha(\by)\vec{v}(\by).\vec{v}(\bx) v_\gamma(\bx) \p^x_\gamma v_\alpha(\bx)\ra  .
\label{Falkovich-relation-2-app}
\end{align}
The term involving the time derivative is proportional to the time derivative of $\la \left( \vec{v}(\bx)\cdot\vec{v}(\by) \right)^2 \ra$
and thus vanishes in the stationary regime.
The force and the velocity are large-scale fields, and thus $\la \vec{f}(\bx)\cdot\vec{v}(\by)\vec{v}(\by)\cdot\vec{v}(\bx)  \ra
\simeq\la \vec{f}\cdot\vec{v}\,\vec{v}^2 \ra$. Using  that the local dissipation operator 
$\epsilon=-\nu v_\alpha(\bx)\p_x^2v_\alpha(\bx)$ is a small-scale field enables one to write 
$\la  v_\alpha(\by) \vec{v}(\by)\cdot\vec{v}(\bx) \nu \Delta^x v_\alpha(\bx)\ra = -\la \epsilon \vec{v}^2\ra/d$.
Finally, by introducing, as in Ref.~\cite{Falkovich10},  $u_\alpha=v_\alpha(\bx)-v_\alpha(0)$
and $V_\alpha=v_\alpha(\bx)+v_\alpha(0)$ that are respectively small- and large-scale fields,  one obtains
\begin{align}
 \la  v_\alpha(\by) \left(\vec{v}(\bx).\vec{v}(\by)\right)^2\ra=&\frac{1}{32}\la u_\alpha \vec{V}^4\ra+
 \frac{1}{32}\la u_\alpha \vec{u}^4\ra \nonumber\\
 &-\frac{1}{16}\la u_\alpha \vec{u}^2\vec{V}^2\ra .
\end{align}
The term $\la u_\alpha \vec{u}^4\ra $ is negligible in the inertial range. The term $\la u_\alpha \vec{V}^4\ra$
 was not present in the other relation Eq.~(\ref{relFalko}) and hence not discussed in Ref.~\cite{Falkovich10}. 
 This function vanishes when $\vert \vec{x}\vert\to 0$ and thus
 $\la u_\alpha \vec{V}^4\ra= F(x^2) x_\alpha$ with $F$ a function that cannot diverge faster than $ \vert \vec{x}\vert^{-1}$.
 Gathering  the previous terms leads to Eq.~(\ref{rel-F-2}).

\end{appendix}

\bibliographystyle{prsty}

\begin{thebibliography}{10}

\bibitem{Falkovich10}
G. Falkovich, I. Fouxon, and Y. Oz, J. Fluid Mech. {\bf 644},  465  (2010).

\bibitem{Smith98}
L. Smith and S. Woodruff, Annu. Rev. Fluid Mech. {\bf 30},  275  (1998).

\bibitem{Zhou10}
Y. Zhou, Phys. Rept. {\bf 488},  1  (2010).

\bibitem{Canet14b}
L. Canet, B. Delamotte, and N. Wschebor, arXiv:1411.7780  (Fully developed
  isotropic turbulence: nonperturbative renormalization group formalism and
  fixed point solution).

\bibitem{dedominicis79}
C. De~Dominicis and P.C. Martin, Phys. Rev. A {\bf 19},  419  (1979).

\bibitem{teodorovich89}
E.~V. Teodorovich, Appl. Math.  Mech. {\bf 53},  340  (1989).

\bibitem{adzhemyan88}
L. Adzhemyan, A.~N. Vasil'ev, and M. Gnatich, Theor. Math. Phys. {\bf 74},  115
   (1988).

\bibitem{adzhemyan94}
L. Adzhemyan, N. Antonov, and T.~L. Kim, Theor. Math. Phys. {\bf 100},  1086
  (1994).

\bibitem{antonov96}
N.~V. Antonov, S.~V. Borisenok, and V. Girina, Theor. Math. Phys. {\bf 106},
  75  (1996).

\bibitem{berera07}
A. Berera and D. Hochberg, Phys. Rev. Lett. {\bf 99},  254501  (2007).

\bibitem{berera09}
A. Berera and D. Hochberg, Nucl. Phys. B {\bf 814},  522  (2009).

\bibitem{Lebedev94}
V.~V. Lebedev and V.~S. L'vov, Phys. Rev. E {\bf 49},  R959  (1994).

\bibitem{kardar86}
M. Kardar, G. Parisi, and Y.-C. Zhang, Phys. Rev. Lett. {\bf 56},  889  (1986).

\bibitem{Canet11a}
L. Canet, H. Chat\'e, B. Delamotte, and N. Wschebor, Phys. Rev. E {\bf 84},
  061128  (2011).

\bibitem{Canet12Err}
L. Canet, H. Chat\'e, B. Delamotte, and N. Wschebor, Phys. Rev. E {\bf 86},
  019904  (2012).

\bibitem{martin73}
P.~C. Martin, E.~D. Siggia, and H.~A. Rose, Phys. Rev. A {\bf 8},  423  (1973).

\bibitem{Janssen76}
H.~K. Janssen, Z. Phys. B {\bf 23},  377  (1976).

\bibitem{DeDominicis76}
C. De~Dominicis, J. Phys. Colloques {\bf 37},  247  (1976).

\bibitem{Note1}
The NS action is strictly invariant under a time-dependent shift of the
  pressure $p(t,\protect \mathaccentV {vec}17E{x})\to p(t,\protect \mathaccentV
  {vec}17E{x})+\epsilon (t)$, as recalled {\protect \it eg.} in \cite
  {altaisky91}.

\bibitem{Note2}
The NS action is strictly invariant under a constant shift of the velocity
  response field, as noted in {\protect \it eg.} \cite {teodorovich89}.

\bibitem{karman38}
T. von K\'arm\'an and L. Howarth, Proc. R. Soc. Lond. A {\bf 164},  192
  (1938).

\bibitem{Frisch95}
U. Frisch, {\em Turbulence: the legacy of A.N. Kolmogorov} (Cambridge
  University Press, Cambridge UK, 1995).

\bibitem{Note3}
Note that the velocity field $v_\alpha $ is unconstrained in the functional
  integral (\ref {fonctgeneratgeneral}). However, since the response pressure
  sector is not renormalized, the incompressibility constraint is in practice
  always satisfied on average at zero external sources.

\bibitem{Kolmogorov41c}
A.N. Kolmogorov, Dokl. Akad. Nauk SSSR {\bf 32},  16  (1941), reprinted in Proc. R. Soc. Lond. A {\bf 434}, 15 (1991). 

\bibitem{altaisky91}
M.~V. Altaisky and S.~S. Moiseev, J. Phys. I France {\bf 1},  1079  (1991).

\end{thebibliography}

\end{document}